# Multilayer graphene, Moiré patterns, grain boundaries and defects identified by scanning tunneling microscopy on the m-plane, non-polar surface of SiC

P. Xu[1], D. Qi[1], J.K. Schoelz[1], J. Thompson[1], P. M, Thibado[1]∗, V.D. Wheeler[2], L.O. Nyakiti[3], R.L. Myers-Ward[2], C.R. Eddy, Jr.[2], D.K. Gaskill[2], M. Neek-Amal[4,#], and F. M. Peeters[4]

[1]Department of Physics, University of Arkansas, Fayetteville, AR, 72701, USA

[2]U.S. Naval Research Laboratory, Washington, DC 20375, USA

[3]Departments of Marine Engineering, Material Science and Engineering, Texas A&M University, College Station TX, 77843 USA

[4]Departement Fysica, Universiteit Antwerpen, Groenenborgerlaan 171, B-2020 Antwerpen, Belgium

[#]Department of Physics, Shahid Rajaee Teacher Training University, Lavizan, Tehran 16788, Iran

**Abstract**

Epitaxial graphene is grown on a non-polar n$^+$ 6H-SiC m-plane substrate and studied using atomic scale scanning tunneling microscopy. Multilayer graphene is found throughout the surface and exhibits rotational disorder. Moiré patterns of different spatial periodicities are found, and we found that as the wavelength increases, so does the amplitude of the modulations. This relationship reveals information about the interplay between the energy required to bend graphene and the interaction energy, i.e. van der Waals energy, with the graphene layer below. Our experiments are supported by theoretical calculations which predict that the membrane topographical amplitude scales with the Moiré pattern wavelength, $L$ as $L^{-1} + \alpha L^{-2}$.

∗ Corresponding author. E-mail: Thibado@uark.edu (P.M. Thibado)

1. **Introduction**

Owing to its unique structure, graphene possesses a number of extraordinary electronic properties including high thermal conductivity,[1] ballistic transport,[2] and ultrahigh electron mobility,[3] making graphene a highly promising candidate for the realization of nano-electronic circuits.[4] The epitaxial growth of graphene on hexagonal SiC surfaces by thermal decomposition is promising for the practical realization of graphene devices, as SiC wafers are already used in existing technology.[5-10] The graphitization of the polar silicon terminated surface starts with the formation of a carbon buffer layer that resembles graphene, but is actually covalently bonded to the SiC substrate. The second carbon layer therefore is the first decoupled graphene plane. Shortly after this discovery, it was determined that the carbon buffer layer can have detrimental effects on the transport properties of the graphene plane due to charge inhomogeneity.[11] This has motivated research efforts to grow graphene on SiC without introducing a carbon buffer layer.

One approach is to grow graphene on one of the non-polar surfaces of SiC, such as the a-plane 6H SiC($11\bar{2}0$) or the m-plane 6H SiC($1\bar{1}00$). In fact, under certain conditions growth on these surfaces does not include the formation of a buffer layer, instead graphene forms directly on the SiC surface.[12] The m-plane is particularly interesting, as the surface is naturally slightly corrugated, which is expressed as alternating mini Si and C facets, with charge transfer between the two facets known to cause n-type doping in the graphene plane.[12, 13]

The presence of the mini-facets is also thought to explain why the growth of graphene is much faster on the m-plane compared to the silicon terminated surface. In fact, Daas et.al., found graphene on the m-plane surface grows at least 8 times faster than on the Si surface.[14] Furthermore, they attribute the formation of nano-crystalline, graphite-like features at higher coverage, to the lack of a hexagonal template, a lower surface energy, and altered step dynamics. Given that graphene is impermeable,[15, 16] any Si atoms that need to escape from below must diffuse along the surface to either a step edge, a defect in the graphene lattice or a grain boundary.[14, 17, 18] The rapid growth process naturally leads to more disorder in the system like twisted bi-layers and grain boundaries, resulting in interesting effects on the electronic properties.[19-27]

Height variation in the graphene influences its physical properties, for example, it sometimes results in a non-uniform strain in graphene which leads to the opening of an energy gap due to the pseudo-magnetic field [28-30]. In fact, earlier density functional theory calculations found in-plane strain in graphene over h-BN opened a gap of 50-60 meV [31]. Several studies on the electronic properties of twisted bilayer graphene and graphene over graphite [32-34] focused on the emergence of new Dirac cones in the density of states (DOS) spectrum.

In this paper, we examine the epitaxial growth of graphene on the non-polar, m-plane of SiC. Using scanning tunneling microscopy (STM) we observe an overall step-terrace structure, which indicates a nominal miscut covering the surface of SiC. A variety of Moiré patterns are found with different wavelength and corrugation amplitude. In some instances, two different Moiré patterns are separated by a sharp line highlighting the differences in amplitude and suggestive of a grain boundary. We observed that the Moiré pattern amplitude decreases as the twisted angle increases, suggesting a competition between the interlayer potential energy and the out-of-plane bending energy of graphene. Here, using ab-initio data [30] and an analytic study, we are able to explain our experimental results of the of height variation, *h*, of graphene Moiré

patterns with wavelength, L. In particular, the following power law dependence was found $h(L) \sim L^{-1} + \alpha L^{-2}$.

## 2. Experimental

### 2.1 Sample preparation

Epitaxial graphene was grown on the m-plane, non-polar surface of an n$^+$ 6H-SiC substrate (Aymont Technology) measuring 16 mm × 16 mm. Growth was carried out in a commercially available hot-wall Aixtron VP508 chemical vapor deposition reactor. Prior to graphene growth, the SiC substrate was etched *in situ* in a 100 mbar H$_2$ ambient at 1560 °C for 50 min. After the H$_2$ etching step, the ambient condition was switched to Ar with a transition time of 2 min. during which pressures varied by ±50% around 100 mbar, this was followed by a temperature ramp to 1620 °C. The graphene synthesis process was then conducted for 15 minutes under a flowing Ar environment of 20 standard liters per min. at 100 mbar, with a substrate temperature still at 1620 °C. After growth, the sample was cooled to 1050°C and Ar annealed for 10 minutes before the final cooling to room temperature and diced into 5 mm × 5 mm coupons. Diamond scribed labels were added to the back side. The post-growth morphology was characterized using atomic force microscopy (AFM) in tapping mode configuration, while the graphene coverage was confirmed using Raman spectroscopy with an excitation wavelength of 532 nm, 8 mW laser as the pump probe having a spot size 0.7 μm. After these characterizations the sample was secured, sealed, and sent to the STM facility.

### 2.1 Sample characterization

Constant-current filled-state STM images were obtained using an Omicron ultrahigh-vacuum (base pressure is 10$^{-10}$ mbar), low temperature model STM operated at room temperature. The sample was mounted with silver paint onto a flat tantalum sample plate and transferred through a load-lock into the STM chamber where it was electrically grounded. STM tips were electrochemically etched from 0.25 mm diameter polycrystalline tungsten wire via a custom double lamella method with an automatic gravity-switch cutoff.[35] After etching, the tips were gently rinsed with distilled water; briefly dipped in a concentrated hydrofluoric acid solution to remove surface oxides, and then transferred into the STM chamber.

## 3. Results and Discussion

A 5 μm × 5 μm AFM image of the sample surface is shown in Fig. 1(a) using a black to orange to white color scale spanning 10 nm. The surface shows a series of steps and terraces that run nearly vertically across the image, and gives the surface a root mean square (RMS) roughness of 0.62 nm and an average step height of 2.5 nm. Confocal Raman spectrometry show a two-phonon intervalley (2D) mode, which confirms the presence of graphene, and the spectrum is shown in Fig 1(b). The nearly Gaussian shaped peak is centered at 2710 cm$^{-1}$ and has a full width half maximum (FWHM) of 60 cm$^{-1}$. The slightly broadened 2D peak indicates that a few layers (3-4) of graphene are present.[36]

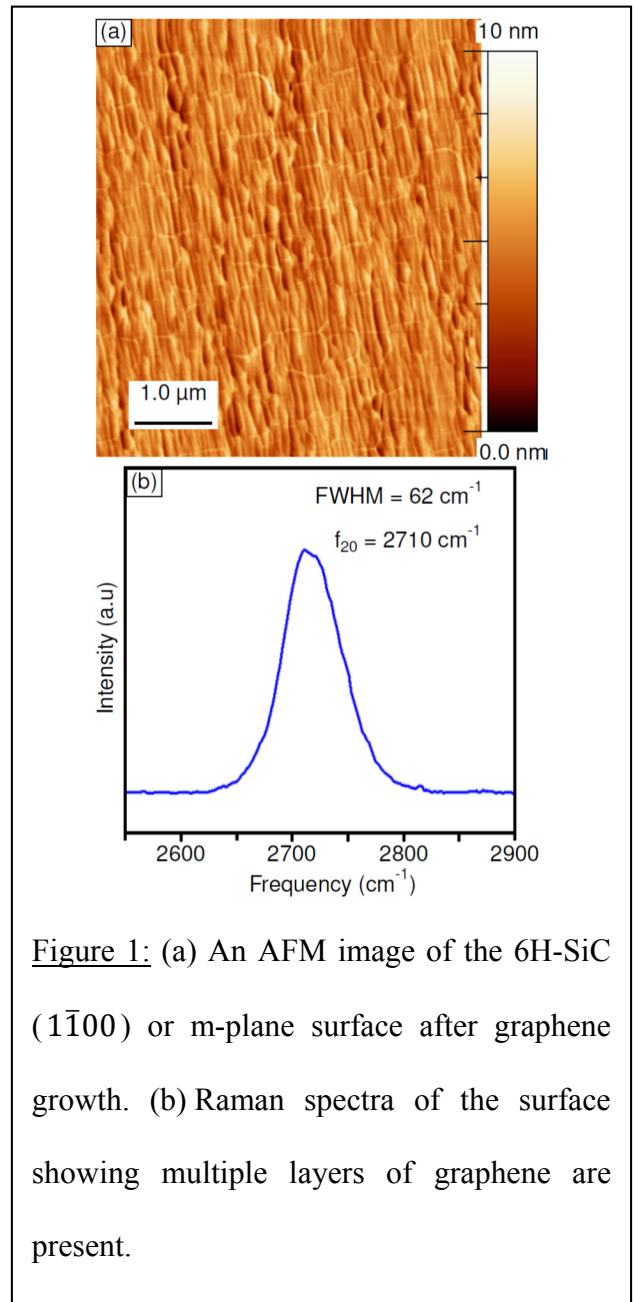

Figure 1: (a) An AFM image of the 6H-SiC $(1\bar{1}00)$ or m-plane surface after graphene growth. (b) Raman spectra of the surface showing multiple layers of graphene are present.

A large scale (1 μm × 1 μm) STM image of the m-plane SiC surface is shown in Fig. 2(a). Similar to the AFM findings, the surface is characterized by a series of steps and terraces that run across the image. A higher magnification STM image measuring 200 nm × 200 nm is shown in Fig. 2(b). Notice the terraces are very uniform and flat throughout, and the step edges are very straight. A diagonal height profile extracted from the image is shown underneath in Fig. 2(c), and reveals the average step height (~2 nm) is about five SiC bilayers.[13] From the line profile, we can estimate that the substrate miscut angle is about 1.5°. Atomic-resolution STM images were obtained on the terraces and a typical example is shown in Fig. 2(d). Here we see a nearly perfect hexagonal structure throughout, with only slight amplitude fluctuations, indicating that this is ideal graphene decoupled from the underlying substrate.

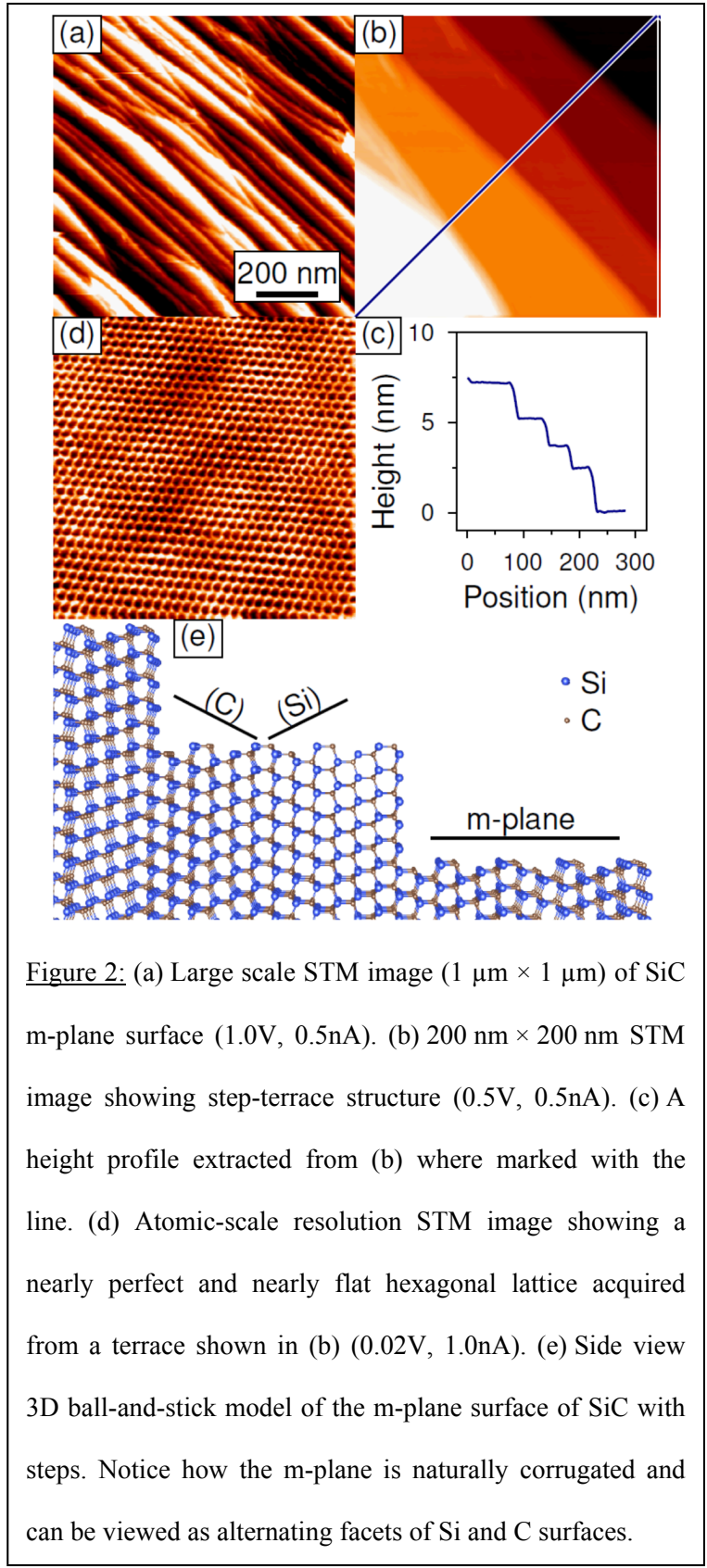

Figure 2: (a) Large scale STM image (1 μm × 1 μm) of SiC m-plane surface (1.0V, 0.5nA). (b) 200 nm × 200 nm STM image showing step-terrace structure (0.5V, 0.5nA). (c) A height profile extracted from (b) where marked with the line. (d) Atomic-scale resolution STM image showing a nearly perfect and nearly flat hexagonal lattice acquired from a terrace shown in (b) (0.02V, 1.0nA). (e) Side view 3D ball-and-stick model of the m-plane surface of SiC with steps. Notice how the m-plane is naturally corrugated and can be viewed as alternating facets of Si and C surfaces.

For comparison, a side view 3D ball-and-stick model of the SiC m-plane surface with large steps is shown in Fig. 2(e). Since the m-plane is non-polar, Si-C bonds can be seen in the very top plane. However, notice there is a slight natural corrugation in the atomic structure, which can be viewed as alternating Si and C facets also marked on the model. Lastly, notice that the front of the step is a Si-terminated polar surface. Up going steps would be the C-terminated polar surface.

Elsewhere on the surface, we see Moiré patterns indicative of multi-layer graphene similar to the one shown in Fig 3(a). A height profile extracted from the marked line location on the image is shown below, and oscillates with amplitude of ~0.1 nm and a wavelength of ~8 nm. Moiré patterns are due to the twisting of two graphene layers that are on top of each other, where the Moiré wavelength, L corresponds a twist angle $\theta = 2\sin^{-1}[a_0/2L]$ where $a_0$=0.142 nm is the C-C bond length. [31] A section of the image is boxed off and shown as a close-up image in Fig. 3(b). Here, we can see the hexagonal lattice of the graphene being modulated by three of the bright features. A further close-up of the boxed off region is shown magnified in Fig. 3(c) and reveals a nearly perfect honeycomb lattice.

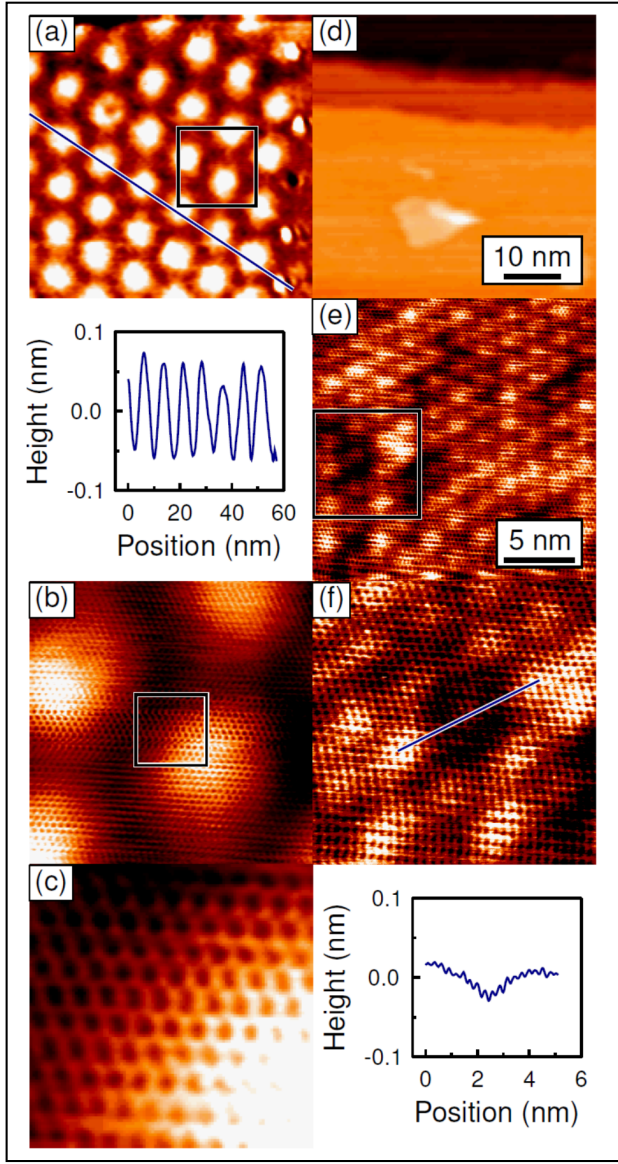

Figure 3: (a) 50 nm × 50 nm STM image showing a Moiré pattern with trigonal symmetry (0.05V, 0.3nA). A height profile extracted from the image shows an average wavelength ~8 nm and amplitude of ~0.1 nm. (b) Higher magnification image from (a) reveals the hexagonal lattice characteristic of graphene (0.05V, 1.0nA). (c) An even higher magnification STM image from (b) (0.05V, 1.0nA). (d) 50 nm × 50 nm STM image showing a terrace structure with an island (0.02V, 1.2nA). (e) A 20 nm × 20 nm STM image taken from the larger terrace shown in (d) (0.05V, 0.5nA). The hexagonal lattice can be resolved nearly everywhere in the image, but is modulated in two ways. The first is a Moiré pattern with a wavelength of ~2 nm, while the second is due to random pits in the surface. (f) A higher magnification STM image from (e) (0.05V, 0.5nA). The hexagonal lattice is resolved nearly everywhere in the image. A slight indentation can be seen near the middle of the image. A height profile extract along the line is shown beneath the STM image.

An STM image measuring 50 nm × 50 nm with the terrace edge in a different direction than before is shown in Fig. 3(d). There is a larger terrace followed by a few steps down to smaller terraces. The larger terrace has a small island in the middle. Notice the edge of the large terrace is not as smooth as before in Fig. 2(b) and that this terrace edge is in a different crystallographic direction about 45º off from the other. A higher magnification image acquired from the large terrace, but away from the small island is shown in Fig. 3(e). Here we see a nearly periodic arrangement of mounds having trigonal symmetry similar to Fig. 3(a), but with a much shorter wavelength and smaller amplitude. In fact, close inspection of some of the tops of mounds reveals the atomic corrugation of the graphene honeycomb structure. Another interesting feature of this image is the numerous yet randomly distributed depressions throughout the image [unlike Fig. 2(d)]. One of these regions, shown inside the box, was reimaged and is displayed in Fig. 3(f). The honeycomb lattice can now be seen throughout, even as the surface ungulates. A height profile extracted from the marked location is displayed beneath the image using the same height scale as before in Fig. 3(a). The darker feature has a depth of only ~0.03 nm, and has graphene throughout. Dual bias imaging gives the same result (not shown), indicating that this is a real topographic depression and not an electronic effect.

The dark regions or pits found in Fig. 3(f), we suspect, are a consequence of the Si diffusion process during graphene growth, as similar pits have been found in ultra-high vacuum grown samples by Sun *et al.,* on the Si (0001) surface.[18] The latter was a very detailed study of the pit formation process where in a wide range of Si growth models combined with STM data they determined that steps at the edge of the terrace play a key role in Si removal. If the step density was too low, for example, the substrate would spontaneously create pits within the boundaries of a given terrace.  This then provided a pathway for the subsurface Si atoms to diffuse out of the substrate and eventually evaporate. In turn, it was also found that by increasing the step density of the substrate via a larger miscut, the number density of pits dropped accordingly. For our surface we have a wide range of terrace widths, and we also have terrace edges along different crystallographic directions. Without a detailed study it is difficult to conclude the role of the terrace edge direction, however, for our limited study it appears that the pits are less likely to form with steps along low index orientations as shown in Fig. 2(b).

The unusual confluence of two Moiré patterns has been found on this surface and is shown in the 110 nm × 50 nm STM image in Fig. 4(a). This is an expanded version of the image shown in Fig. 3(a). The left hand side of the image contains a clear Moiré pattern, while the right half side is mostly flat. The nearly vertical line in the middle marks the unusually abrupt transition from one region to the next. The height profile along the border is shown beneath the image in Fig. 4(b). The line profile highlights four features, separated by ~8 nm distance including two peaks, a single pit, and a third peak. The peaks are unusually high at ~0.25 nm, while the pit has a depth of ~0.1 nm. This boundary marks where a Moiré pattern having an average wavelength of ~8 nm stops ($\theta$~1.76º, $\theta$ represents twisted angle.) and another Moiré pattern with an average wavelength of ~2 nm starts ($\theta$~7.05º). The 8 nm Moiré pattern abruptly ends in the middle of a high point in the pattern, and the strain in the graphene lattice must also be very high at the boundary, in order for it to have such a high amplitude. In fact, the observation of similar boundary structures has been previously reported in the literature.[37, 38] Even more interesting, is that the 2 nm Moiré pattern has such a small amplitude that with this color scale it appears flat.

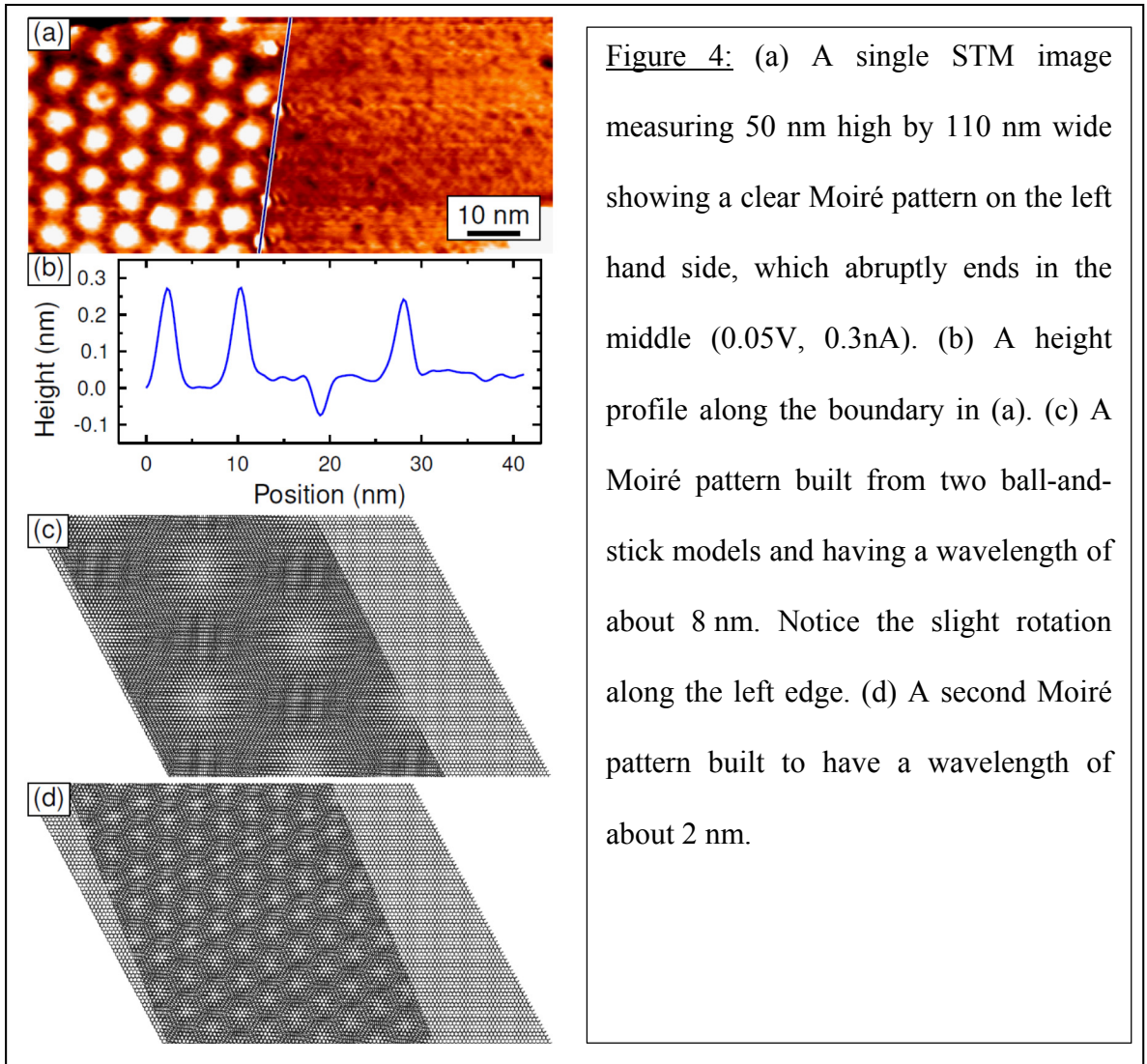

Figure 4: (a) A single STM image measuring 50 nm high by 110 nm wide showing a clear Moiré pattern on the left hand side, which abruptly ends in the middle (0.05V, 0.3nA). (b) A height profile along the boundary in (a). (c) A Moiré pattern built from two ball-and-stick models and having a wavelength of about 8 nm. Notice the slight rotation along the left edge. (d) A second Moiré pattern built to have a wavelength of about 2 nm.

Two sets of Moiré patterns were generated to fit the average wavelengths found in Fig. 4(a). An angular rotation of ~1.8° is shown in Fig. 4(c) and this result is the best match with the data on the left-hand side of Fig 4(a). An angular rotation of ~7° is shown in Fig. 4(d) and gives the best match with the data on the right-hand side of Fig. 4(a) [which is similar to the data shown in Fig. 3(e)]. Even though the Moiré patterns allow us to match the wavelength of the data in the STM images, they do not provide any information about the expected amplitude of the Moiré pattern. To address this, we collected together a variety of our STM images of various multi-layer graphene moiré patterns (not all are shown) from a-plane and m-plane SiC substrates

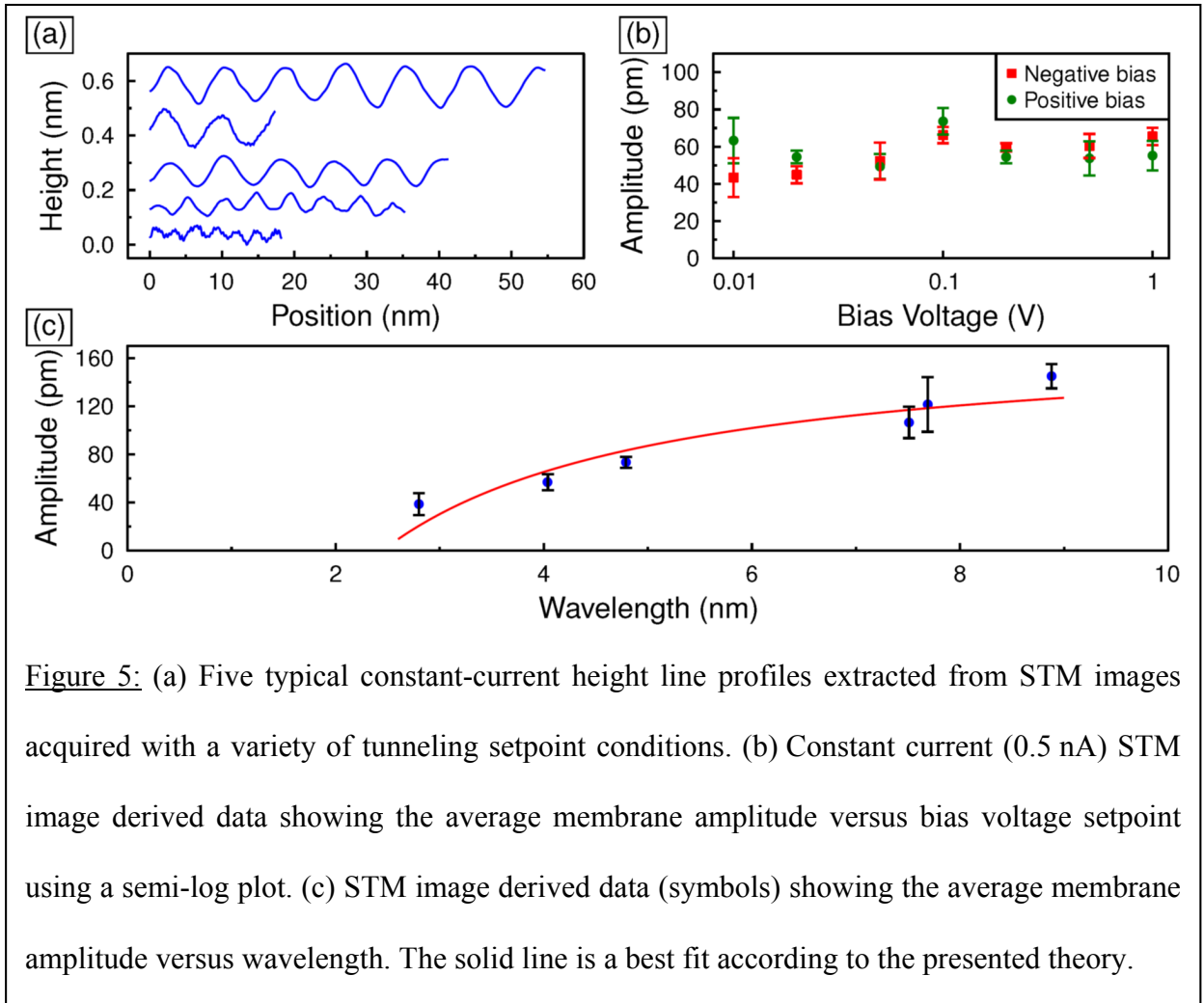

Figure 5: (a) Five typical constant-current height line profiles extracted from STM images acquired with a variety of tunneling setpoint conditions. (b) Constant current (0.5 nA) STM image derived data showing the average membrane amplitude versus bias voltage setpoint using a semi-log plot. (c) STM image derived data (symbols) showing the average membrane amplitude versus wavelength. The solid line is a best fit according to the presented theory.

grown under similar conditions. Five typical line profiles extracted from these STM images and having varying wavelength and amplitude are shown in Fig. 5(a). The line profiles are ordered from top to bottom based on decreasing amplitude. Notice the lowest line profile has, superimposed on it, an even smaller amplitude and higher frequency signal. This is the electronic corrugation of the carbon atoms, and it is worth pointing out how small the electronic amplitude is when compared to the membrane corrugation. Also, the membrane corrugation persists while imaging the moiré pattern through a range of normal bias voltage settings (±0.05 to ±1.00 V) and tunneling current setpoints (0.05 to 1.00 nA). For example, when a moiré pattern with a wavelength of ~4 nm is repeatedly imaged while incrementally altering the bias voltage from

±0.01 V to ±1.0 V with a tunneling current setpoint of 0.5 nA we see the amplitude variation shown with a semi-log plot in Fig. 5(b). Within the error bars the membrane amplitude is relatively unchanged. However, the observed amplitude using STM is the total amplitude containing two sources, i.e. $h = h_m + h_e$, where $h_m$ is the membrane corrugation (membrane topographical amplitude due to movement of the nuclear positions) and $h_e$ is due to the electronic corrugation of the individual atoms. For flat graphene or graphite $h_m$ is zero. By calculating the height change from the top of the electronic corrugation at the top of the Moiré pattern to the top of the electronic corrugation at the bottom of the Moiré pattern, we obtain only the membrane amplitude. Note, for our STM study on the a-plane SiC surface we would sometimes observe the top layer moving under the influence of the bias voltage of the STM tip,[39] but this did not happen on the m-plane and we believe the impact on our ability to measure the moiré pattern wavelength and amplitude is within the error bars we report here. A plot showing the membrane amplitude as a function of L is shown in Fig. 5(c). Even though it is possible that the electronic amplitude is slightly different at the crest of the membrane compared to the trough of the membrane, we believe this is within the error bars of our results. Also, unlike monolayer graphene grown on Ir(111),[40] for twisted graphene on graphene/SiC we do not see any significant height changes in the moiré pattern as we vary the STM tunneling condition. Note that the bright feature in all Moiré patterns is where we have local AA stacking of graphene (i.e., one benzene ring stacked direct over another benzene ring). We can understand this by realizing that in between two adjacent AA stacks there is a low energy AB (i.e., Bernal) stacked region. Since carbon atoms in an AA stack have higher energy as compared to the one for the AB stack we expect larger amplitude in AA stacked region[41], i.e. the AB stacked planes are much closer together as compared to AA stacked planes. Because of the infinitesimal height variation we assume that between AA and AB stacked regions we have local interlayer distances which can be approximated as

$$d_a(\theta) = d_{AB} + \beta\theta \tag{1}$$

where ab-initio calculations [30] gives

$$\beta = \frac{3|d_{AA} - d_{AB}|}{\pi} \tag{2}$$

Furthermore the local binding energy -$E_b(\theta)$- (due to the weak vdW interaction between the layers) has the well-known $1/d^{4.2}$ dependence [42, 43] therefore we write

$$E_b(\theta) = E_{AB}\left(\frac{d_{AB}}{d_a}\right)^{4.2} \tag{3}$$

where the $E_{AB} \sim -100$ meV/atom [36]. Substituting Eq. (1) in Eq. (3) leads us to write $E_b(\theta) = E_{AB}\left(\frac{1}{1+\frac{\beta}{d_{AB}}\theta}\right)^{4.2}$. The binding energy is competing with the bending energy (elastic energy, $E_e$) and in mechanical equilibrium we must have $E_b(\theta) = E_e(\theta)$. On the other hand using Monge representation and reduced length units, the elastic energy can be roughly estimated as

$$E_e(\theta) \sim \left(\frac{a_0}{\theta}\right)^2 d_a^2(\theta) \tag{4}$$

Comparing Eq. (4) and Eq. (3), and using $\theta \cong \frac{\sqrt{3}a_0}{L}$ after a straightforward algebra, gives us the height variation with Moiré wavelength

$$h_m(L) \sim \frac{1}{L} - \frac{1}{L^2}\frac{(2\sqrt{3}a_0\beta)}{\pi d_{AB}} \sim L^{-1} + \alpha L^{-2} \tag{5}$$

Therefore, the membrane corrugations vary non-linearly with $L$ and consequently the total amplitude will have the same behavior. In Fig. 5(c) we show the variation of our measured $h_m$ with respect to $L$ (symbols) and the corresponding results from the above theory by the solid line. Since the Moiré pattern combines regions of AA and AB stacked graphene, the lowest energy configuration of the membrane should involve oscillations between two different equilibrium heights. However, transitioning between the two heights will increase the bending energy of the graphene membrane. For the shorter wavelength Moiré patterns, the increase in the bending energy gives rise to smaller amplitudes. While increasing the distance between successive AA stacks lowers the bending energy of the graphene membrane and naturally gives rise to larger amplitudes.

## 4. Conclusion

We successfully grew epitaxial graphene on the non-polar, m-plane surface of an $n^+$ 6H-SiC substrate. STM images reveal a variety of surface topographies, includes atomically flat terraces with sharp step edges, and pitted terraces with rough edges along a different crystallographic orientation. This may be an indication of subsurface Si diffusion anisotropy. We also observed a large variety of Moiré patterns with several different twist angles, and found a sharp boundary between two different Moiré patterns. We have also discovered a non-linear relationship between the amplitude of the Moiré pattern and the wavelength of the Moiré pattern. We argue that this relationship is due to a competition in energy between the inter-planar bonding and the top layer bending.

**Acknowledgements**

P.X. and P.M.T. gratefully acknowledge the financial support of ONR under grant N00014-10-1-0181 and NSF under grant DMR-0855358. L.O.N. acknowledges the support of American Society for Engineering Education and Naval Research Laboratory Postdoctoral Fellow Program. Work at the U.S. Naval Research Laboratory is supported by the Office of Naval Research. This work was supported by the Flemish Science Foundation (FWO-Vl), the Methusalem Foundation of the Flemish Government, and the EUROgraphene project CONGRAN. M.N.-A was supported by the EU-Marie Curie IIF postdoc Fellowship 299855.